\documentclass[ reprint,amsmath,amssymb,aps,pra,]{revtex4-1}

\usepackage{graphicx}
\usepackage{dcolumn}
\usepackage{bm}
\usepackage{mathrsfs}
\usepackage{ulem}
\usepackage{color}

\newcommand{\ket}[1]{\left| #1 \right\rangle}
\newcommand{\overlap}[2]{\left\langle #1 | #2 \right\rangle}
\newcommand{\expect}[2]{\left\langle #2 \left| #1 \right| #2 \right\rangle}
\newcommand{\average}[1]{\left\langle #1 \right\rangle}

\begin{document}

\preprint{APS/123-QED}

\title{Probabilistic Eigensolver with a Trapped-Ion Quantum Processor}

\author{Jing-Ning Zhang$^1$}
\author{I\~nigo Arrazola$^2$}
\author{Jorge Casanova$^2$}
\author{Lucas Lamata$^2$}
\author{Kihwan Kim$^1$}
\author{Enrique Solano$^{2,3,4}$}
\affiliation{%
$^1$Center for Quantum Information, Institute for Interdisciplinary Information Sciences, Tsinghua University, Beijing 100084, People's Republic of China\\
$^2$Department of Physical Chemistry, University of the Basque Country UPV/EHU, Apdo. 644, 48080 Bilbao, Spain\\
$^3$IKERBASQUE, Basque Foundation for Science, Maria Diaz de Haro 3, 48013 Bilbao, Spain\\
$^4$Department of Physics, Shanghai University, Shanghai 200444, China
}

\date{\today}

\begin{abstract}
Quantum simulation of complex quantum systems and their properties often requires the ability to prepare initial states in an eigenstate of the Hamiltonian to be simulated. In addition, to compute the eigenvalues of a Hamiltonian is in general a non-trivial problem. Here, we propose a hybrid quantum-classical probabilistic method to compute eigenvalues and prepare eigenstates of Hamiltonians which are simulatable with a trapped-ion quantum processor. 

\end{abstract}

\maketitle

\section{Introduction}\label{sec:introduction}

Simulating complex quantum systems is known to be an intractable task if accomplished with classical computing resources. This is because the dimension of the Hilbert space needed to describe a quantum system increases exponentially with its number of constituents. Quantum simulation~\cite{Feynman82}, i.e. the use of a fully controlled quantum system that simulates the dynamics of another one, was originally proposed to overcome this problem, and has led to plenty of developments in different quantum platforms~\cite{Georgescu14}. On the other hand, quantum computing is a more general approach that allows not only to simulate quantum dynamics, but also to solve systems of linear equations~\cite{Harrow09}, linear differential equations~\cite{Berry17,Xin18,Arrazola18}, or the eigenvalue problem of complex Hamiltonians~\cite{Abrams99,Nielsen00}, with an exponential speedup. Regarding the latter, it is important to remark that the knowledge of the energy spectrum of a system is crucial in different fields such as quantum chemistry and condensed matter physics. 

The {\it phase estimation algorithm}~\cite{Abrams99} is a prominent method for finding eigenvalues and preparing eigenstates of target Hamiltonians, however other approaches have been also designed. Among them, we can mention the use of adiabatic evolutions~\cite{Aspuru05}, algorithmic cooling~\cite{Xu14} or variational quantum eigensolvers~\cite{Peruzzo14, Yung14, McClean16}. The latter are designed to prepare the ground state (or excited state~\cite{Shen17}) of a complex Hamiltonian, which interestingly combine a quantum processor and a classical optimization algorithm. The variational eigensolvers have been implemented experimentally in state-of-the-art quantum platforms like photonics~\cite{Peruzzo14}, superconducting circuits~\cite{Kandala17}, or trapped ions~\cite{Hempel18}, and their accuracy in eigenstates preparation, as well as in the computing of the associated eigenvalues, highly depends on the flexibility of the ansatz~\cite{Malley16,Shen17,Kandala17}. In this respect, having a more complex and flexible input state, or ansatz, implies that more variational parameters have to be optimized simultaneously, which could be haunted by the presence of local minima. In addition, the preparation of the ansatz may involve non-trivial entanglement operations among the available quantum registers. In this manner, it is important to explore alternative algorithms to be applied in near-future quantum processors~\cite{Debnath16}.

In this article, we propose a hybrid classical-quantum eigensolver that uses projective measurements on a single ancilla qubit to probabilistically prepare the eigenstates of target Hamiltonians in trapped ions. Unlike the quantum variational solver or adiabatic state preparation, in our method the only requirement for the initial state is to have a non-negligible overlap with the desired target eigenstate. In Sec.~\ref{sec:general}, we first introduce the general method for preparing an arbitrary target eigenstate, given that the overlap between the target and the initial state is nonzero. Afterwards, we present how the method can be combined with a classical optimization algorithm, resulting in a significant reduction on the number of steps required for the preparation, while maintaining acceptable success probability. In Sec.~\ref{sec:toolkit}, we present the toolkit that trapped-ion quantum simulators offer to implement our method. Finally, in Sec.~\ref{sec:examples}, we present some models in which our probabilistic eigenstate preparation could be applied using a trapped-ion quantum processor, and demonstrate its performance with numerical simulations.

\section{General framework}\label{sec:general}
Consider a target quantum system described by the Hamiltonian $\hat H$ such that 
\begin{eqnarray}
\hat H\ket{j}=E_j\ket{j}.
\end{eqnarray}
Here, $\ket{j}$ ($E_j$) is the $j$-th eigenstate (eigenenergy) and the energy spectrum $\left\{E_j\right\}$ is bounded from below and sorted in ascending order. To prepare one of the eigenstates of $\hat H$, we consider a quantum system isomorphic to the target system and an additional ancillary qubit. The evolution of the whole system is described by a unitary operator acting on the product Hilbert space ${\mathscr H}_{\rm S}\otimes{\mathscr H}_{\rm A}$, with ${\mathscr H}_{\rm S}$ (${\mathscr H}_{\rm A}$) being the Hilbert space of the target system (the ancilla qubit).

For preparing the $j$-th eigenstate $\ket{j}$, our method requires the  knowledge of all eigenenergies below $E_j$.  As, in general, this is not the case, we have to start with the ground state preparation. In this manner, this first step can be considered as a cooling protocol. To this end, we apply the following unitary operator on the whole system
\begin{eqnarray}\label{eq:U_tau}
\hat W_\gamma(\tau) = \exp\left[-i\left(\hat H_{\rm S}+\gamma\right)\hat\sigma_{\rm A}^{\rm x}\tau\right].
\end{eqnarray}
Here, $\hat\sigma_{\rm A}^{\rm x}$ is the x Pauli matrix acting on the ancillary qubit and the parameters $\gamma$ and $\tau$ are real numbers to be determined. $\hat W_\gamma(\tau)$ can be viewed as an evolution operator of the whole system with $\tau$ being the effective evolution time. We use $\gamma$ to shift the energy spectrum of $\hat H_{\rm S}$ such that $E_0+\gamma\geq0$ (i.e. the shifted spectrum is always positive), where $E_0$ is the ground state energy of the target system.

Our protocol is an iterative method where each repetition involves, first, initializing the ancillary qubit to a reference state $\ket{0}_{\rm A}$ such that $\hat\sigma^z_{\rm A}\ket{0}_{\rm A}=-\ket{0}_{\rm A}$, and second, applying $\hat W_\gamma(\tau)$. In the $k$-th iteration we will obtain a state $\ket{\Psi_{k}}$, which is
\begin{eqnarray}
\ket{\Psi_{k}}&=&\hat W_\gamma(\tau)\ket{\psi_{k-1}}\ket{0}_{\rm A}\nonumber\\
&=&\hat{\mathcal C}_\gamma(\tau)\ket{\psi_{k-1}}\ket{0}_{\rm A}-i\hat{\mathcal S}_\gamma(\tau)\ket{\psi_{k-1}}\ket{1}_{\rm A}.
\end{eqnarray}
Here, $\ket{\psi_{k-1}}$ is the normalized state vector in ${\mathscr H}_{\rm S}$ that results from the $k-1$ iteration, and the non-unitary operators are defined as $\hat{\mathcal C}_\gamma(\tau)=\cos\left[\left(\hat H+\gamma\right)\tau\right]$ and $\hat{\mathcal S}_\gamma(\tau)=\sin\left[\left(\hat H+\gamma\right)\tau\right]$. Finally, we perform a projective measurement on the ancillary qubit and make post-selection on the measurement results to select certain the output state in ${\mathcal H}_{\rm S}$. The two possible normalized output states, conditioned on the measurement results, are $\ket{\psi_k}_0=\hat{\mathcal C}_\gamma(\tau)/{\sqrt{P^{0}_{k}}}\ket{\psi_{k-1}}$
and $\ket{\psi_k}_1=\hat{\mathcal S}_\gamma(\tau)/{\sqrt{P^{1}_k}}\ket{\psi_{k-1}}$, and their probabilities are $P^{0}_k =\sum_j\left|\overlap{j}{\psi_{k-1}}\right|^2\cos^2\left[\left(E_j+\gamma\right)\tau\right]$ and $P_k^{1}=\sum_j\left|\overlap{j}{\psi_{k-1}}\right|^2\sin^2\left[\left(E_j+\gamma\right)\tau\right]$, respectively. We would like to point out that, although we use a pure input state  $\ket{\psi_{k-1}}$ to illustrate the process,  our method can be straightforwardly applied to mixed input states without any modification.

Now, if we compare the average energy of the two possible output states with the average energy of the input state, we find that, for positive and small $\tau$, the following relation holds (See Appendix A)
\begin{eqnarray}\label{eq:average_energy}
\average{\hat H}^{(0)}_{k}\leq\average{\hat H}_{k-1}^{(0)}<\average{\hat H}^{(1)}_{k}.
\end{eqnarray}
Here $\average{\cdot}_k^{(i)}\equiv\expect{\hat H}{\psi_k^{(i)}}$ and the superscript $i = 0, 1$ denotes the outcome of the projective measurement on the ancillary qubit. In other words, the average energy of the system is lowered compared to the one of the input state, i.e. the post-selected state from the previous round, if we successfully project the ancillary qubit in $\ket{0}_{\rm A}$.  Here, it is noteworthy to comment that the probability of projecting on the reference ancilla state $\ket{0}_{\rm A}$ (i.e. on the state that leads to an effective average energy reduction) is close to one for a sufficiently small $\tau$ (note that $P_k^0=1$ if $\tau=0$). An intuitive way to understand this energy reduction for small $\tau$ is provided when studying the transformation of the state. If $\ket{\psi_{k-1}}=\sum_{j}c^{k-1}_{j}\ket{j}$ is the system input state before the $k$-th iteration, the output state after the application of $\hat W_\gamma(\tau)$ and post-selection of $\ket{0}_{\rm A}$ will be $\ket{\psi_{k}}=\ket{\psi_k}_0\equiv\sum_{j}c_j^{k}\ket{j}$, where the probability distribution of the state have changed as
\begin{eqnarray}\label{eq:prob_dist}
\left|c_j^k\right|^2=\frac{1}{P_{k}^{0}} \cos^2{[(E_j+\gamma)\tau]}\left|c_j^{k-1}\right|^2.
\end{eqnarray}
As $\cos^2(\cdot)$ is a monotonic decreasing function in the vicinity of $\tau=0$, the probability amplitudes of those eigenstates with lower eigenenergies $E_j$ will be enhanced (note that a larger value of $E_j$ implies a lower value for the corresponding $\cos^2(\cdot)$ function) compared to those with higher eigenenergies, leading to the decrease of the average energy, an effective {\it cooling} effect.

After the $k$-th iteration, we have to measure the average energy of the resulting state $\bar E^{k}\equiv \average{\hat H}^{(0)}_{k}$, and compare it with the average energy of the previous stage $\bar E^{k-1}$. If $ |\bar E^{k-1}-\bar E^{k}|<\epsilon$ holds (with $\epsilon$ being a small positive value), we conclude that that the ground-state within a certain precision $\epsilon$ has been reached and stop the protocol. If the state does not converge at the $k$-th iteration, saying $ |\bar E^{k-1}-\bar E^{k}|>\epsilon$, the output state $\ket{\psi_{k}}_0$ will be sent through another iteration for further reducing its energy.

\begin{figure}
  \includegraphics[width=0.45\textwidth]{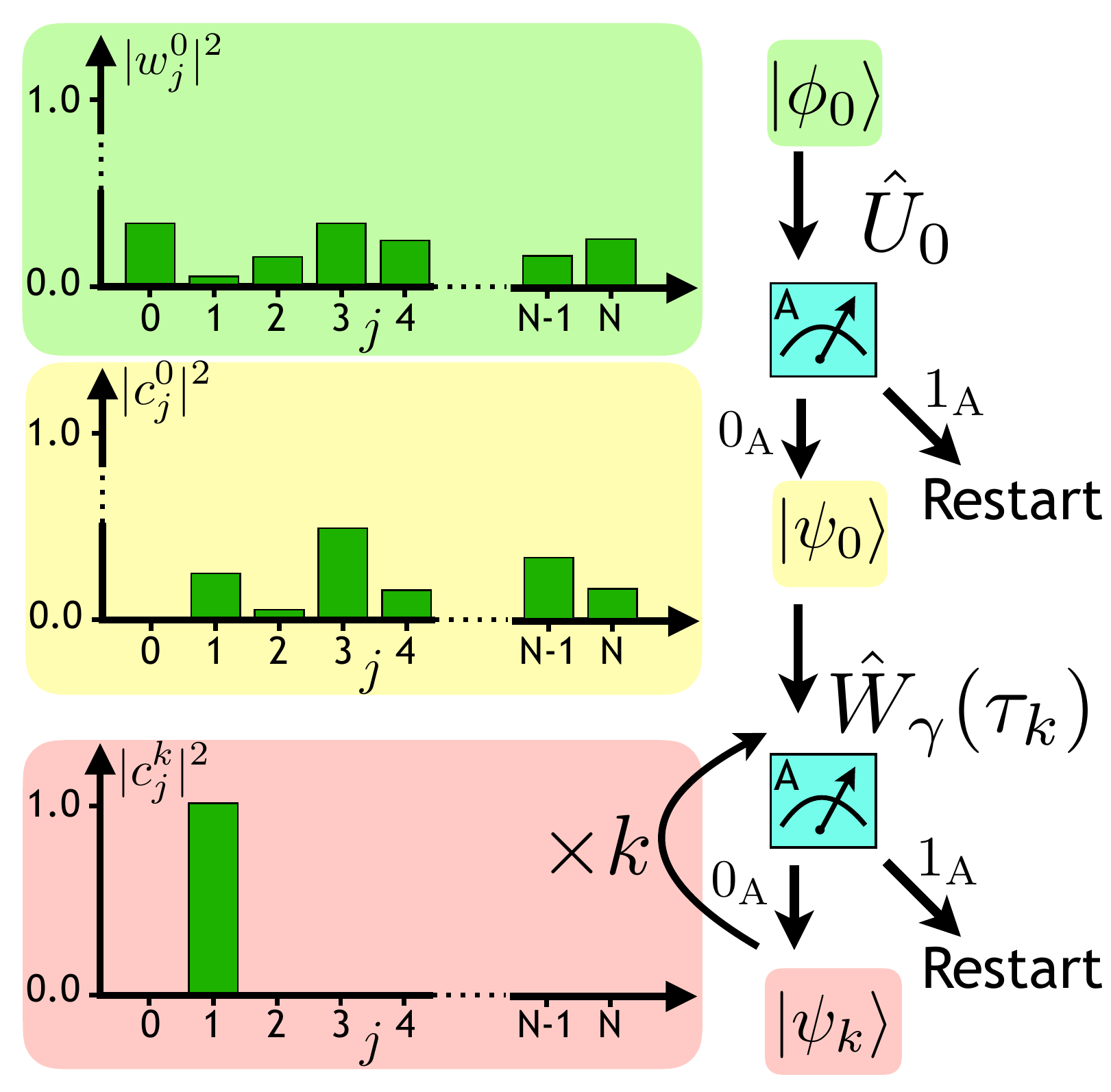}\\
  \caption{Scheme for preparing the first excited state of an $N$ dimensional Hamiltonian. First, we apply $\hat{U}_0$ and measure the ancilla qubit. Then we use post-selection to remove the ground-state contribution from the initial state $\ket{\phi_0}$. After that, we apply the cooling unitary $\hat{W}_\gamma(\tau_k)$ and subsequent ancilla measurement $k$ times, until we reach convergence.}\label{fig:Fig1}
\end{figure}

An important feature of our protocol is that if the initial state $\ket{\psi_0}$ has no overlap with the desired eigenstate, it is not possible to prepare it. It can be seen from Eq.~(\ref{eq:prob_dist}) that if the $(k-1)$-th input state has no overlap with a given eigenstate $\ket{j}$ (i.e. $c_{j}^{k-1}=0$), the resulting $k$-th state will also share the same condition $c_{j}^{k}=0$. Although this may seem a disadvantage, it can actually be used to prepare  excited eigenstates. To this end, we start from an arbitrary initial state $\ket{\phi_0}$ and apply the unitary operation 
\begin{eqnarray}
\hat U_{s}=\exp\left[-i\left(\frac{\pi}{2 E_{s}}\right)\hat H\hat\sigma_{\rm A}^{\rm x}\right],
\end{eqnarray}
and measure the ancilla qubit. The action of the unitary operator $\hat U_{s}$ on a general state $\ket{\phi}$ is to conditionally produce a state $\ket{\psi^{(0)}}$ where the contribution of the basis state $\ket{s}$ goes to zero ($\langle s |\psi^{(0)}\rangle=0$), as long as the ancilla qubit is projected on $\ket{0}_{\rm A}$. As a result, the output state after post-selection does not have any overlap with the eigenstate $\ket{s}$. For example, to prepare the first excited state of $\hat H$ starting from an arbitrary initial state $\ket{\phi_0}$, one starts with applying the unitary $\hat U_{0}$ and projecting the ancilla on the $\ket{0}_{\rm A}$ state. If the projection succeeds, one will obtain a state $\ket{\psi_0}$ that has zero overlap with the ground state of $\hat H$, which serves as input state for the {\it cooling stage}. Then, one  applies the previously described protocol (See Fig.~\ref{fig:Fig1}), until the state converges to the lowest energy eigenstate that has a non-zero overlap with the initial state, which in this case is the first excited state.

To prepare an arbitrary eigenstate $\ket{j}$ starting from $\ket{\phi_0}$, one first removes any overlap with lower-energy eigenstates, by sequentially applying $\hat U_{j'}$ and the following projective measurement with $j' = 0, 1, \ldots, j-1$. If all of the post-selection operations succeed, the resulting state is sent to the {\it cooling stage}, which would probabilistically converge to the desired eigenstate $\ket{j}$. Note that once a post-selection operation fails, the whole procedure should be started all over again.

We consider now two different variants for the {\it cooling stage}, namely the {\it fixed-step} and {\it variational} approaches.

{\it The fixed-step approach} treats $\tau$ as a small fixed parameter. The operational procedure is as follows: After each implementation of $\hat W_\gamma(\tau)$, we perform a projective measurement on the ancilla qubit. We continue with the next iteration if the outcome is $\ket{0}_{\rm A}$, otherwise we restart the whole process. At the $k$-th stage of the process, we estimate the average energy of the output state $\bar E^{k}$. The process stops if $|\bar E^{k-1}-\bar E^{k}| \leq \epsilon$, where the precision $\epsilon$ is a positive small value. Otherwise, we continue the protocol by moving to the $(k+1)$-th stage. 

\begin{figure*}
\centering
\includegraphics[scale = 0.5]{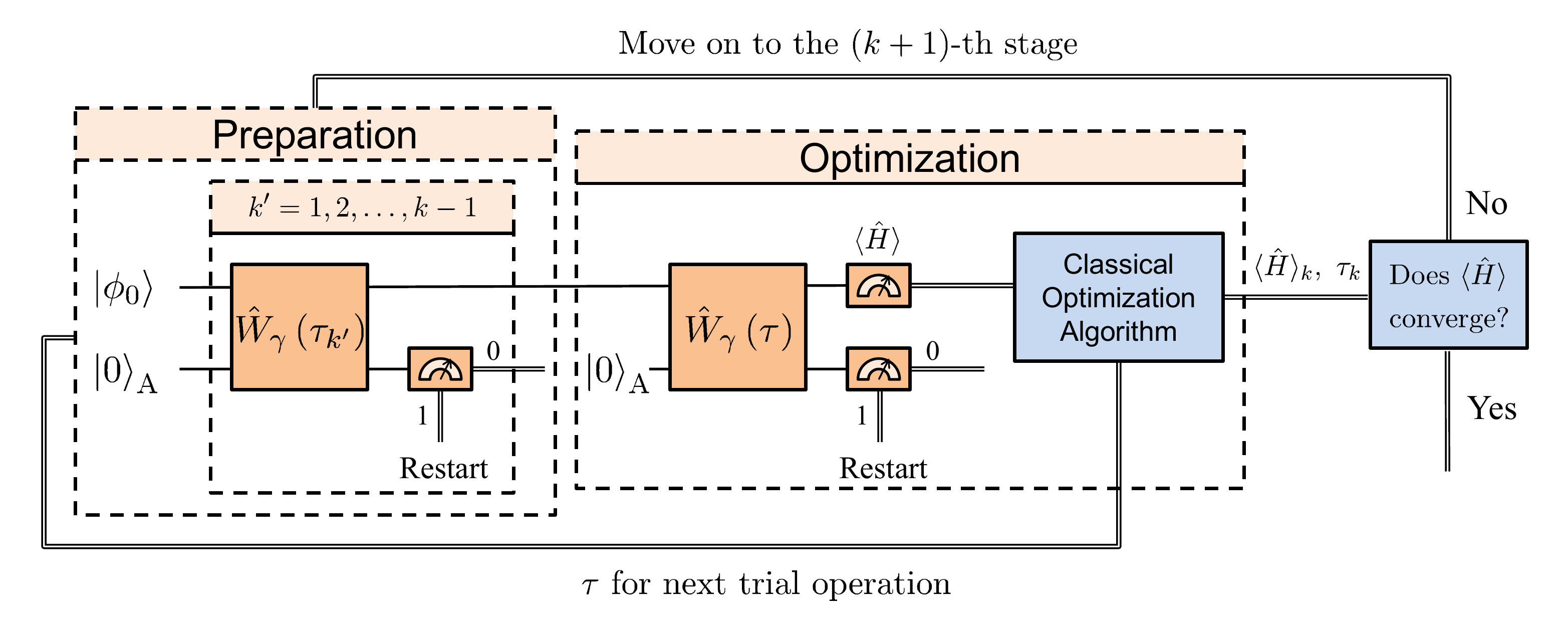}
\caption{Scheme for obtaining the optimal value of $\tau$ for the $k$-th stage in the variational approach to the probabilistic eigensolver. First, prepare the state at the $k-1$-th stage by applying the effective cooling $k-1$ times with optimized parameters $\left\{\tau_1, \tau_2, \ldots, \tau_{k-1}\right\}$. Then, search for the optimal value $\tau_k$ with the help of a classical optimization algorithm. Specifically, the quantum simulator estimates the average energy and feeds it into the classical algorithm~\cite{Bounded}, which determines whether the optimal value is reached or alternatively the next trial value for $\tau$.}
\label{fig:variational}
\end{figure*}

In the {\it variational approach}, the value of the parameter $\tau$ is optimized in each stage such that it minimizes the average energy. This hybrid optimization is done by feeding the average energy obtained from the quantum platform to a classical optimization algorithm, which provides new trial value for $\tau$ to the quantum platform based on the given information (see Fig.~\ref{fig:variational}). After several optimization steps (trial steps), the classical algorithm will converge to $\tau_k$, the optimal value of $\tau$ at the $k$-th stage, and the quantum platform will estimate the average energy $\bar E^k$. As before, the process stops if $|\bar E^{k-1}-\bar E^{k}| \leq \epsilon$. At the end, we obtain a set of optimized values $\left\{\tau_1, \tau_2, \ldots, \tau_{k}\right\}$. Compared to the {\it fixed-step} approach, the {\it variational} approach allows the probabilistic preparation of the desired eigenstate with a substantial reduction on the circuit depth, i.e. with a qualitatively less amount of gates and post-selections.

Both the {\it fixed-step} and the {\it variational} approaches are probabilistic because we need to restart the whole procedure depending on the results of the projective measurements on the ancilla qubit, and provide an estimation of the eigenenergy of the target eigenstate within the preset precision $\epsilon$. However, the variational protocol additionally provides a recipe for shallow-depth probabilistic preparation of the target state, described by the set of optimized parameters $\{\tau_1,\ldots,\tau_k\}$.   The overall success probability $P_{\rm suc}$ of both protocols is ultimately  bounded by the overlap between the initial state and the desired eigenstate, i.e, $P_{\rm suc} \leq |\langle\psi_0|j'\rangle|^2$. On the other hand, to maximize $P_{\rm suc}$ for the preparation of eigenstate $|j'\rangle$, one should choose the value of $\gamma$ to be the closest possible to $-E_{j'}$. However, it is possible that we do not know the exact value of $E_{j'}$, in which case one can always choose $\gamma$ to be minus the lower bound of the norm of the Hamiltonian, which can be analytically approximated in general.

Finally, we comment that the unitary operator $\hat W_\gamma(\tau)$ required in our method can be decomposed as $\hat W_\gamma(\tau)\equiv\hat W(\tau)R_{\rm x}^{\rm A}\left(2\gamma\tau\right)$, with $\hat W(\tau)=\exp\left(-\hat H\hat\sigma_{\rm A}^{\rm x}\tau\right)$ and $R_\alpha^{\rm A}\left(\theta\right)\equiv\exp\left(-\frac{i\theta}{2}\hat\sigma^\alpha_{\rm A}\right)$, $\alpha = {\rm x, y, z}$, being the single qubit rotation acting on the ancilla qubit. Now, the $\hat W(\tau)$ operator that entangles the system and the ancilla qubit, can be trotterized into a set of basic operations~\cite{Lloyd96}. More specifically, suppose that the system Hamiltonian is composed of $M$ components $\hat H_m$ that don't commute with each other, namely $\hat H = \sum_{m=1}^M\hat H_m$. Then, the unitary evolution in Eq.~(\ref{eq:U_tau}) can be approximately written using the second-order Trotter-Suzuki expansion,
\begin{eqnarray}
\hat W(\tau)&=&\left[\hat W_1\left(\frac{\tau}{2r}\right)\ldots\hat W_{M-1}\left(\frac{\tau}{2r}\right)\hat W_M\left(\frac{\tau}{r}\right)\right.\nonumber\\
&&\left.\times\hat W_{M-1}\left(\frac{\tau}{2r}\right)\ldots\hat W_1\left(\frac{\tau}{2r}\right)\right]^r+{\mathcal O}\left(\frac{\tau^3}{r^2}\right),
\end{eqnarray}
where $r$ is the number of Trotter steps and $\hat W_m(\tau) = \exp\left(-i\hat H_m\hat\sigma_{\rm A}^{\rm x}\tau\right)$ with $m= 1,2,\ldots M$. Note that by increasing the number of Trotter steps, arbitrary precision can be achieved in principle. 

\section{Toolkit in trapped-ion platforms}\label{sec:toolkit}

Trapped ions have been demonstrated to be suitable for digital~\cite{Lanyon11} and analog~\cite{Zhang17,Safari17} quantum simulations, having access not only to the manipulation of the ion-qubits, but also to the coherent control of their collective vibrational modes~\cite{Toyoda15}. The digital-analog approaches which combines both resources could enhance the computational power of the quantum platform for solving hard problems~\cite{Shen14}, for example, non-trivial many-body models or even problems in quantum field theories~\cite{Casanova11}.

We envision four types of primitive system-ancilla couplings in the toolkit of trapped-ion systems, including: 1) $\hat\sigma_i^{\rm z}\hat\sigma_{\rm A}^{\rm x}$, 2) $\hat a_m^\dag\hat a_m\hat\sigma_{\rm A}^{\rm x}$, 3) $\hat\sigma_{i_1}^{\rm x}\hat\sigma_{i_2}^{\rm x}\hat\sigma_{\rm A}^{\rm x}$, 4) $\left(\hat a_m^\dag+\hat a_m\right)\hat\sigma_i^{\rm x}\hat\sigma_{\rm A}^{\rm x}$. Using these interactions and the measurement on the ancilla, the protocol described in the previous section could be implemented in trapped-ion systems for a wide variety of models. 

The free Hamiltonian $\hat H_0$ of an array of $N$ ions trapped in a linear Paul trap is written as follows,
\begin{eqnarray}
\hat H_0 = \sum_{n=1}^N\frac{\hbar\omega_{\rm 0}}{2}\hat\sigma^{\rm z}_n+\sum_{m = 1}^N\hbar\omega_m\left(\hat a_m^\dag\hat a_m+\frac{1}{2}\right),
\end{eqnarray}
where $\hat\sigma^{\rm x, y, z}_n$ are the Pauli matrices acting on the $n$-th ion with $\omega_{\rm 0}$ being the frequency splitting of the two involved internal levels, and $\hat a_m$ ($\hat a_m^\dag$) is the annihilation (creation) operator of the $m$-th collective motional mode with the mode frequency $\omega_m$. For a better readability, from now on we will omit the subscript $n$ ($m$) of $\hat\sigma_n^{\rm x,y,z}$ ($\hat a_m$) when there is no ambiguity.

In the following, we provide a brief explanation of how to engineer the four types of system-ancilla couplings in a trapped-ion quantum simulator:

First, spin-spin couplings can be implemented by the M\o lmer-S\o rensen (MS) type of interaction, which takes the form of $\sigma_i^{\rm x}\sigma_j^{\rm x}$ and can be engineered between any pair of ions embedded in a linear chain of ions~\cite{Leung18}. With the appropriate single qubit rotations on the first qubit, we can extend the coupling to the form $\hat\sigma^{\mathbf n}\sigma_A^{\rm x}$, that would correspond to a simulated local spin-1/2 operator $\hat H_{\rm S}\propto\hat\sigma_{\mathbf n}$, where ${\mathbf n}=\left(n_{\rm x}, n_{\rm y}, n_{\rm z}\right)^{\rm T}$ is a unit vector.

Second, the $\hat a^\dag\hat a\hat\sigma_{\rm A}^{\rm x}$ interaction term involving the $m$-th mode could be realized with a detuned red-sideband coupling, for which the Hamiltonian is written, in the rotating frame with respect $\hat H_0$ and after the rotating wave approximation, as
\begin{eqnarray}
\hat H_{\rm I}&=&\frac{\hbar\Omega_{\rm A}}{2}\left[\left(\hat\sigma^+e^{i\Delta t}+\hat\sigma^-e^{-i\Delta t}\right)\right]\nonumber\\
&&+\frac{i\hbar\eta\Omega_{\rm A}}{2}\left(\hat a\hat\sigma^+e^{-i\delta t}-\hat a^\dag\hat\sigma^-e^{i\delta t}\right),
\end{eqnarray}
where $\Omega_{\rm A}$ is the ancilla Rabi frequency, $\eta\equiv\eta_m^{\rm A} = \delta k\sqrt{\frac{\hbar}{2M\omega_m}}$ is the Lamb-Dicke (LD) parameter associated to the $m$-th mode and the ancilla qubit, with $\delta k$ being the transferred wave-vector of the laser and $M$ being the mass of the ion, and $\Delta$ and $\delta$ are the detunings to the carrier and the first red-sideband transitions, respectively. In the dispersive regime, i.e. $\Omega_{\rm A}\ll\Delta$ and $\eta\Omega_{\rm A}\ll\delta$, the effective Hamiltonian can be written as follows~\cite{Solano05},
\begin{eqnarray}
\hat H_{\rm eff} = \frac{\hbar\Omega_{\rm A}^2}{4\Delta}\hat\sigma^{\rm z}+\frac{\hbar\eta^2\Omega_{\rm A}^2}{4\delta}\left(\hat a^\dag\hat a+\frac{1}{2}\right)\hat\sigma^{\rm z},
\end{eqnarray}
where the AC-Stark shift term can be absorbed into the free Hamiltonian, and the term that depends on the bosonic operators can be transformed into $\hat a^\dag\hat a\hat\sigma_{\rm A}^{\rm x}$ by appropriate single-qubit rotations on the ancilla.

\begin{figure}
  \includegraphics[width=0.45\textwidth]{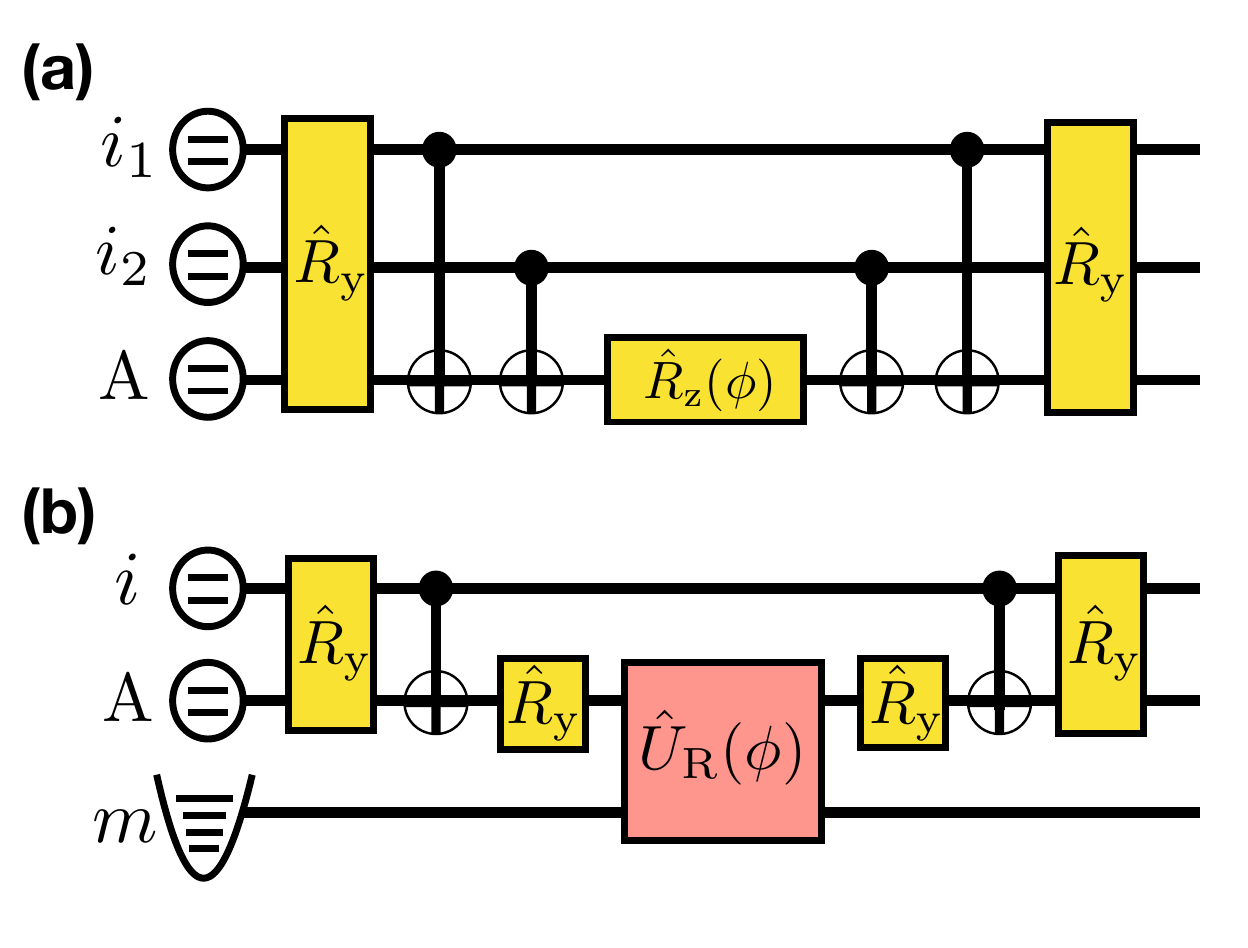}\\
  \caption{Quantum circuits for the trapped-ion toolkit. (a) Scheme for implementing the unitary $\exp{[-\frac{i\phi}{2}\hat\sigma_{i_1}^{\rm x}\hat\sigma_{i_2}^{\rm x}\hat\sigma_{\rm A}^{\rm x}]}$ using C-NOT gates and single-qubit rotations. Here, $\hat R_{\rm y}$ represents the action of a single-qubit rotation $\hat R_{\rm y}\equiv\exp{(i\pi \hat\sigma_{\rm y}/4)}$. (b) Scheme for implementing the unitary operator $\exp{[-\frac{i\phi}{2}\left(\hat a+\hat a^\dag\right)\hat\sigma_i^{\rm x}\hat\sigma_{\rm A}^{\rm x}]}$ using C-NOT gates, single-qubit gates and the analog block $\hat U_{\rm R}(\phi)$ involving the $m$-th motional mode.}\label{fig:Fig2}
\end{figure}

The third term,  $\hat\sigma^{\rm z}_1\hat\sigma^{\rm z}_2\hat\sigma^{\rm z}_3$, can be implemented by combining the controlled-not (C-NOT) gates~\cite{Nielsen00} or the MS gates~\cite{Muller11} with single-qubit z-rotations. Fig.~\ref{fig:Fig2}(a) shows a recipe to construct this nonlocal spin operator with C-NOT gates.

Finally, similarly to the previous quantum circuit, we show that the dynamics governed by $\left({\hat a}^\dagger+\hat a\right)\hat\sigma_i^{\rm x}\hat\sigma_{\rm A}^{\rm x}$ can be implemented by combining C-NOT gates (or MS gates~\cite{Casanova11_2}) with analog blocks that involve the dipolar coupling between the $m$-th motional mode with the ancilla qubit. In Fig.~\ref{fig:Fig2}(b) we show the complete circuit to achieve the desired interaction. The central part of the circuit is a unitary operator produced by a simulated dipolar coupling that can be implemented by simultaneously driving the red-sideband and blue-sideband transitions of the ancilla qubit~\cite{Solano02}. The associated evolution operator $\hat U_{\rm R}(\phi)$ in the rotating frame defined by the free Hamiltonian $\hat H_0$ is written as follows,
\begin{eqnarray}
\hat U_{\rm R}(\phi) = \exp\left[-\frac{i\phi}{2}\left(\hat a+\hat a^\dag\right)\hat\sigma_{\rm A}^{\rm x}\right],
\end{eqnarray}
where $\phi$ is proportional to the Rabi frequency $\Omega_{\rm A}$, the Lamb-Dicke parameter $\eta_m^{\rm A}$, and the interaction time. Another qubit will be involved by a pair of C-NOT gates and all the other single-qubit rotations are used for changing the basis of the Pauli matrices.

With the described toolkit available on trapped-ion quantum platforms, the probabilistic eigensolver can be extended to a wide range of models involving both qubits and bosonic modes. The latter include spin-spin interaction models like the Ising, XY or Heisenberg models~\cite{Arrazola16}, the quantum Rabi model~\cite{Pedernales15,Puebla17,Puebla16}, the Dicke models~\cite{Aedo18}, or second-quantization Hamiltonians like the Holstein model~\cite{Mezzacapo12}. Also, more general methods like the ones for computing $n$-time correlation functions~\cite{Pedernales14} or to simulate dissipative processes~\cite{DiCandia15} could benefit from the presented eigensolver as the latter can be used to prepare arbitrary eigenstates. In the next section, we numerically investigate the performance of the probabilistic eigensolver with several examples, including a single harmonic oscillator, the quantum Rabi model, and the Hubbard model.

\section{Examples}\label{sec:examples}

\subsection{Harmonic Oscillator}\label{subsec:ho}
One of the simplest and most important models of quantum mechanics is the quantum harmonic oscillator, which gives a mathematical description of various physical phenomena, including mechanical oscillators in harmonic potentials and electromagnetic fields~\cite{Sakurai11}. The Hamiltonian of a single harmonic oscillator is written as follows,
\begin{eqnarray}\label{eq:h_harmonic}
\hat H=\hbar\omega_{\rm h.o.}\hat a^\dag\hat a,
\end{eqnarray}
where $\hat a$ ($\hat a^\dag$) is the annihilation (creation) operator of the mode with the mode angular frequency $\omega_{\rm h.o.}$. 

The Hamiltonian in Eq. (\ref{eq:h_harmonic}) provides an equal-spacing spectrum with an infinite number of levels. One could think on a simple example of the protocol presented in Sec.~\ref{sec:general}, where the application of a unitary of the form $\hat W(\tau)=\exp{[-i\tau\hat a^\dag \hat a \hat \sigma_{\rm A}^{\rm x}]}$  (see Sec.~\ref{sec:toolkit}) in a single ion, combined with measurements on the ancilla, could lead to a probabilistic preparation of the motional ground state. Note that the ground state energy for a harmonic oscillator is positive, thus $\gamma$ is set to zero. Starting from, for example, a thermal state, each time the ancilla is measured to be in the state $\ket{0}_{\rm A}$ after applying $\hat W(\tau)$, the motional state is effectively ``cooled down" (i.e. the average energy is lowered).

\begin{figure}
  \includegraphics[width=0.4\textwidth]{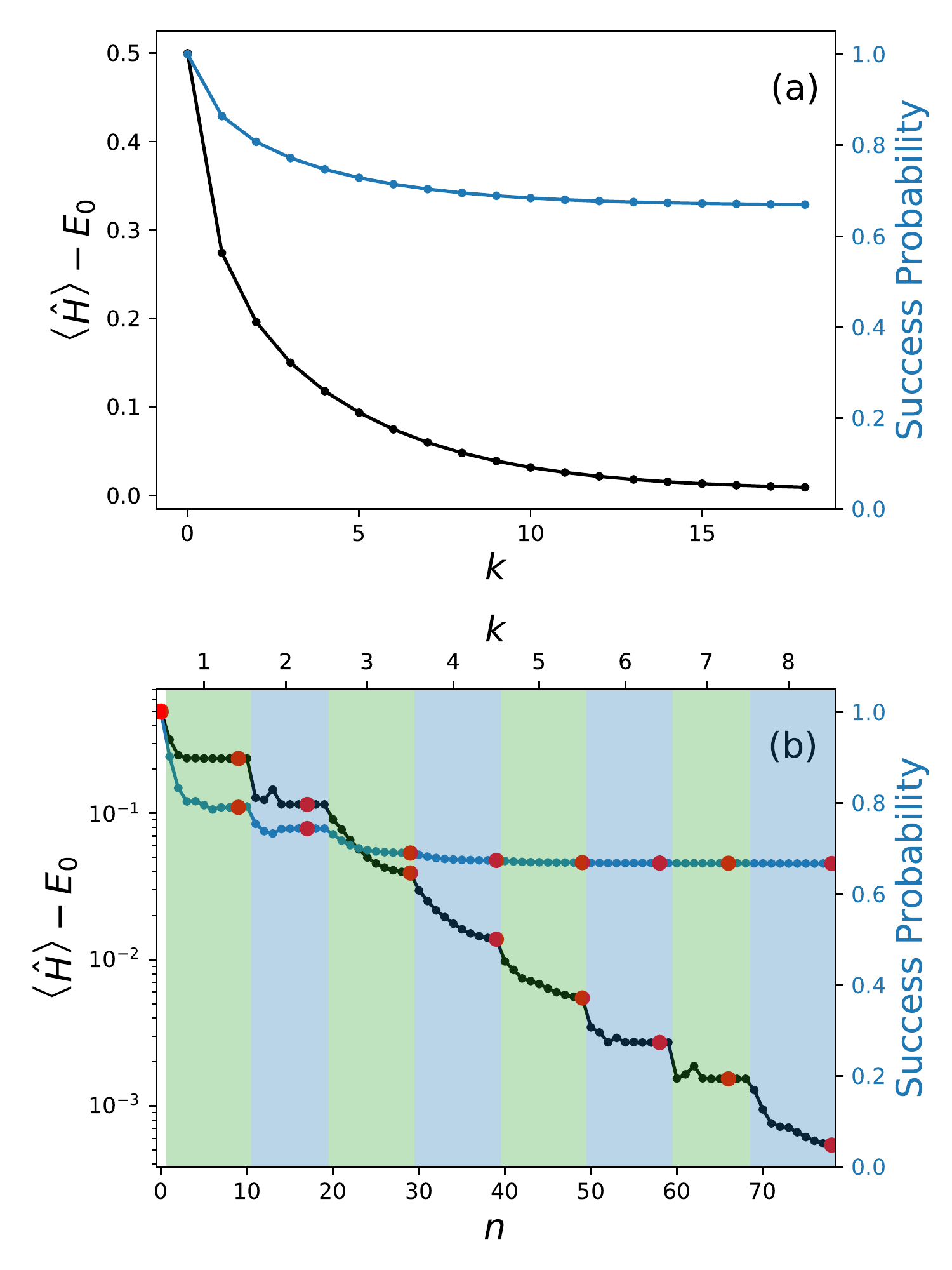}\\
  \caption{Probabilistic ground-state preparation of a single harmonic mode. (a) Average energy $\average{\hat H}$ (black dots) and the overall success probability $P_{\rm suc}(k)$ (blue dots) as functions of the number of iterations $k$ for the fixed-step protocol. The normalized step size $\tau$ is fixed to be $0.3$. (b) Average energy $\average{\hat H}$ (red dots over the black curve) and the overall success probability $P_{\rm suc}(k)$ (red dots over the blue curve) as functions of the number of iterations $k$ for the optimized variational protocol. Black dots represent results for all the different trial steps that are needed until reaching convergence for the optimum $\tau_k$ which are $\{0.8487, 0.5044, 0.9919, 0.9919, 0.9910, 0.7430, 0.4194, 0.9881\}$.}\label{fig:cooling_harmonic}
\end{figure}

In Fig.~\ref{fig:cooling_harmonic}, we show the results for the probabilistic ground-state preparation of a harmonic oscillator with $\omega_{\rm h.o.}=1$. The initial state is naturally chosen as a thermal equilibrium state characterized by the thermal average number $\bar n_{\rm th} = 0.5$.  Both the fixed-step cooling with $\tau = 0.3$ (Fig.~\ref{fig:cooling_harmonic} (a)) and the variational cooling (Fig.~\ref{fig:cooling_harmonic} (b)) show a monotonically decreasing average energy $\average{\hat H}$ with respect to the number of iterations $k$. The overall success probability $P_{\rm suc}(k)$ up to the $k$-th stage is defined as 
\begin{eqnarray}
P_{\rm suc}(k)= \prod_{k' = 1}^kP_{k'}^0,
\end{eqnarray}
where $P_{k'}^0$ is the probability of projecting the ancilla to $\ket{0}_{\rm A}$ at the $k'$-th iteration. It is shown in Fig.~\ref{fig:cooling_harmonic} that both in the fixed-step and variational cases, $P_{\rm suc}(k)$ decreases towards a saturate value around $0.65$. In this example, the fixed-step and the variational protocols reach convergence in $18$ and $8$ steps respectively, where a precision of $\epsilon=10^{-3}$ is pursued. In Fig.~\ref{fig:cooling_harmonic} (b), it is shown that the total number of trial steps (black dots) needed to obtain the optimized variational protocol are around $80$ ($\sim10$ trials per step). However, once that you have the set of optimized values $\{\tau_k\}$, the variational approach requires fewer steps than the fixed-step approach, which reduces the circuit depth and alleviates the requirement for long-term quantum coherence.

\subsection{Quantum Rabi model}
The quantum Rabi model~\cite{Rabi36} describes the dipolar interaction between a magnetic dipole and an oscillating magnetic field. Recently, its quantum simulation has been realized experimentally with highly tunable parameters in many physical platforms, including superconducting circuits~\cite{Braumuller17,Langford17} and trapped ions~\cite{Lv18}. The Hamiltonian of the quantum Rabi model can be separated into two non-commuting parts, $\hat H = \hat H_1+\hat H_2$, with
\begin{eqnarray}\label{eq:h_rabi}
\hat H_1 &=& \frac{\hbar\omega_0}{2}\hat\sigma^{\rm z}+\hbar\omega\hat a^\dag\hat a,\nonumber\\
\hat H_2 &=& \hbar g\left(\hat a+\hat a^\dag\right)\hat\sigma^{\rm x},
\end{eqnarray}
where $\hat\sigma^{\rm x, z}$ are the Pauli matrices acting on the two-level system with the energy splitting $\omega_0$, $\hat a$ ($\hat a^\dag$) is the annihilation (creation) operator of the harmonic mode with frequency $\omega$, and $g$ is the coupling strength of the dipolar interaction.

We implement $\hat W(\tau)$ through the second-order Trotter-Suzuki expansion,
\begin{eqnarray}
\hat W_\gamma(\tau) = \left[\hat W_1\left(\frac{\delta\tau}{2}\right)\hat W_2\left(\delta\tau\right)\hat W_1\left(\frac{\delta\tau}{2}\right)\right]^r,
\end{eqnarray}
where $\delta\tau=\tau/r$ is the size of the Trotter step and $\hat W_j(\tau)=\exp\left(-i\hat H_j\hat\sigma_{\rm A}^{\rm x}\tau\right)$ with $j = 1$ and $2$.  $\hat W_1$ could be implemented combining the first and second interactions described in Sec.~\ref{sec:toolkit}, while $\hat W_2$ requires the implementation of the fourth interaction. Overall, the simulation requires two qubits and a motional mode. Note that depending on the step size $\tau$, more than one Trotter step may be needed.

\begin{figure}
  \includegraphics[width=0.45\textwidth]{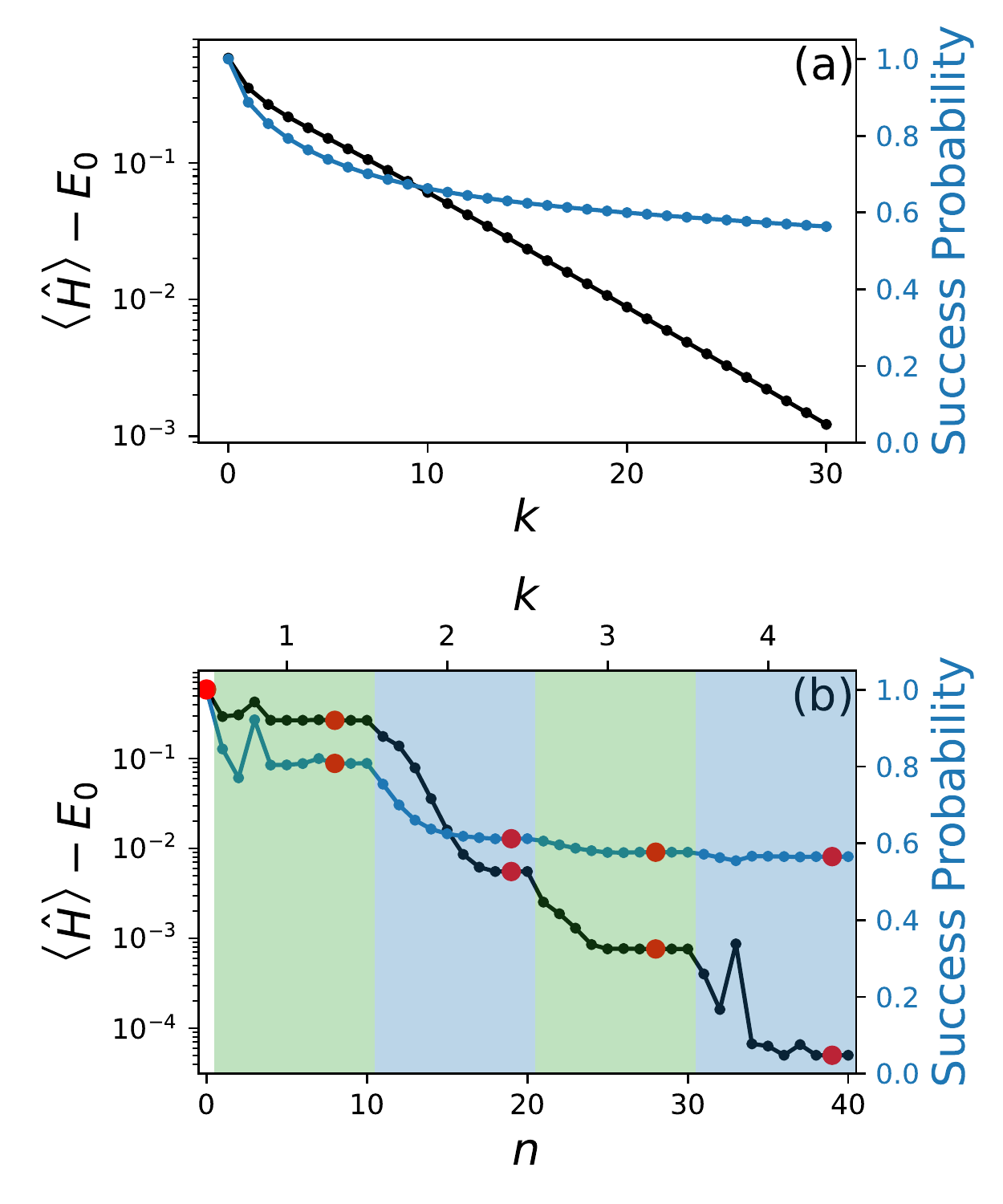}\\
  \caption{Probabilistic ground-state preparation of the quantum Rabi model with $\omega_0 = 1.2$, $\omega = 0.8$ and $g = 1.0$. The energy is in units of $g$ and the number of stages is denoted by $k$. The number of Trotter steps is chosen as $r = 3$. (a) Fixed-step probabilistic state preparation. The step size is chosen as $\tau_{\rm fix}=0.3$. (b) Variational probabilistic state preparation. The total number of trial steps is denoted by $n$ (black dots). The trial steps for different stages are discriminated by different background colours. Red dots represent the energy and overall probability for different stages, using the optimized $\tau$ values, $\{0.4762, 0.9839, 0.9032, 0.5575\}$.}\label{fig:cooling_Rabi}
\end{figure}

Fig.~\ref{fig:cooling_Rabi} shows the results for the ground state preparation of the quantum Rabi model in the deep strong coupling regime~\cite{Casanova10}. We choose the state $\ket{\downarrow}\ket{0}$ as the initial state, for it can be easily prepared by optical pumping and sideband cooling. Comparing both cases we clearly see how the variational approach offers a faster route to the ground state preparation, with an overall success probability around $60\%$, very similar to the fixed-step approach.

\subsection{Hubbard Model}

The Hubbard model~\cite{Hubbard63} is one of the most important models in solid-state physics. In spite of the simplicity in form, it describes the quantum phase transition between superconducting and Mott-insulator phases. The analytical solution has not been available for arbitrary dimensions and the exact numerical treatment is believed to be hard for classical computers because of the sign problem.

In order to illustrate the performance of our protocol in a system with only qubits, we consider the one-dimensional Hubbard model with $L$ sites and open boundary conditions, whose Hamiltonian is written as follows,
\begin{eqnarray}
\hat H&=&-t\sum_{i=1}^{L-1}\sum_\sigma\left(\hat c_{i,\sigma}^\dag\hat c_{i+1,\sigma}+{\rm H.c.}\right)\nonumber\\
&&+U\sum_{i=1}^L\hat c_{i,\uparrow}^\dag\hat c_{i,\uparrow}\hat c_{i,\downarrow}^\dag\hat c_{i,\downarrow},
\end{eqnarray} 
where $t$ and $U$ are the nearest-neighbour hopping strength and the on-site interaction strength, respectively. This model can be mapped to a neighrest-neighbour spin chain of $2N$ spins through the Jordan-Wigner transformation~\cite{Jordan28}, which establishes a mapping between a fermionic operator and a set of spin operators. We choose the one-dimensional Hubbard model as an example, while the extension to 2D and 3D models can be done in trapped ions using nonlocal interactions~\cite{Casanova11}. The operator in Eq.~(\ref{eq:U_tau}) can be constructed using the three body spin operators described in Sec.~\ref{sec:toolkit} and the Suzuki-Trotter expansion. For the simulations, we choose $t=1$ and $U=2$ if not explicitly specified otherwise.

\begin{figure}
  \includegraphics[width=0.45\textwidth]{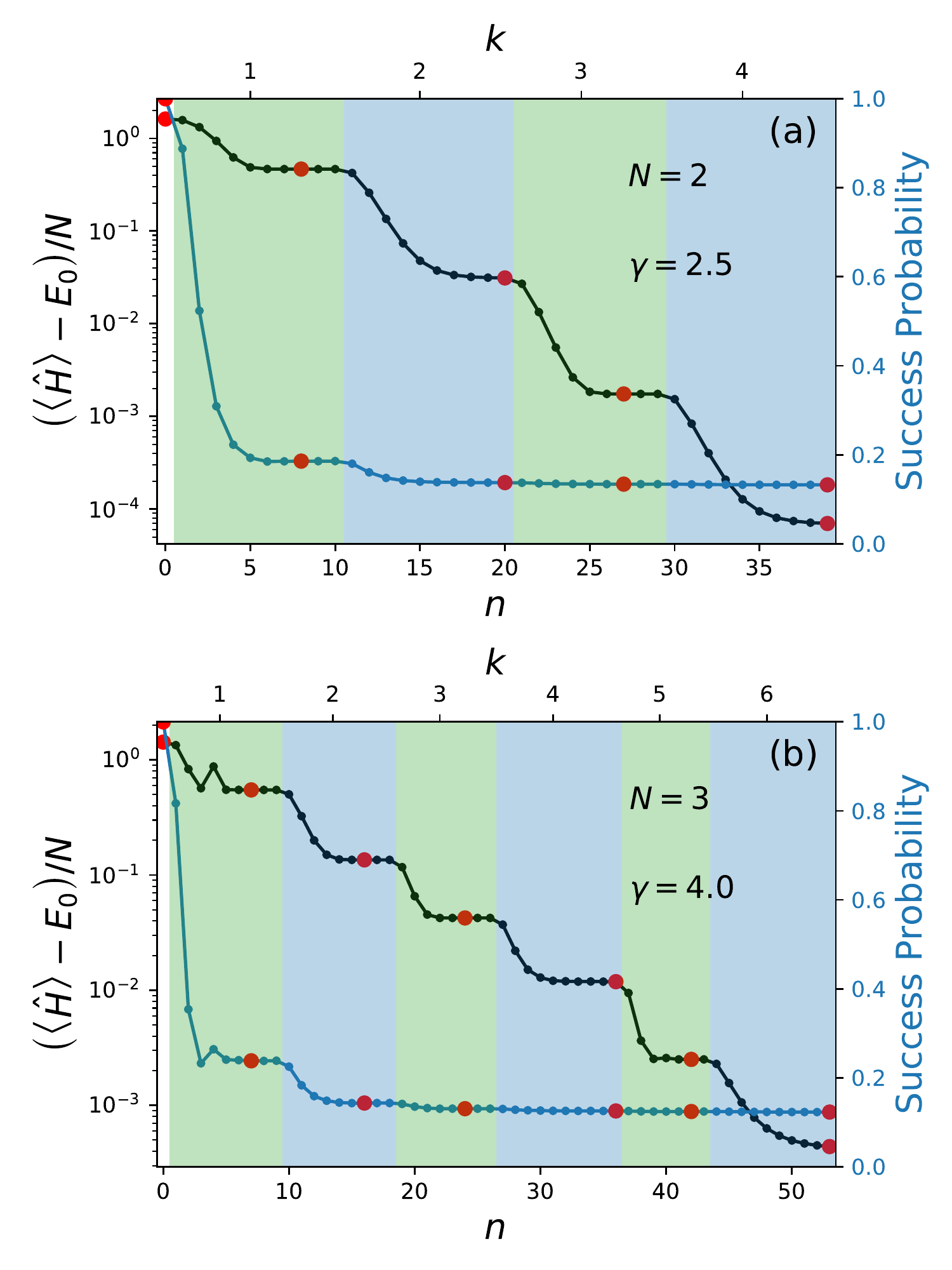}\\
  \caption{Probabilistic ground state preparation of 2-site (a) and 3-site (b) Hubbard models. The black dots indicate the average energy of the trial states after successfully projecting the ancilla qubit to the reference state $\ket{0}_{\rm A}$, while the blue dots represent the overall success probability. Red dots represent the average energy and overall success probability after the $k$-th stage, which is done by applying the $\hat W_{\gamma}(\tau)$ unitary with the optimized $\tau_k$, using $r=3$ Trotter steps. The optimized taus are $\{0.3692, 0.3959, 0.3684, 0.3959\}$ and $\{0.2694, 0.3661, 0.3247, 0.3973, 0.2726, 0.3959\}$ for the 2-site and 3-site models, respectively.}\label{fig:cooling_Hubbard}
\end{figure}

Fig.~\ref{fig:cooling_Hubbard} shows the numerical simulations for the ground state preparation of the 2-site and 3-site Hubbard models, using the variational approach for probabilistic state preparation. As initial states, we choose $\ket{\uparrow\downarrow\uparrow\downarrow}$ and $\ket{\downarrow\uparrow\downarrow\downarrow\uparrow\downarrow}$ for the 2-site and 3-site model respectively. In addition, the $\hat W_{\gamma}(\tau)$ operation is implemented using the symmetric Trotter expansion with $r=3$ Trotter steps.  The results show that with less than $10$ steps, one can obtain the ground state for both models. The overall success probability is in this case around $15\%$ which is still an acceptable value.

\section{Conclusions}\label{sec:conclusions}

We have presented a probabilistic eigensolver which is capable of preparing arbitrary eigenstates of Hamiltonians that are implementable with a trapped-ion quantum processor. The method, applicable to a digital or digital-analog quantum simulator, requires control operations and measurements on an extra ancilla qubit. Moreover, we provide a recipe to enhance the performance of the probabilistic eigenstate preparation by means of a hybrid classical-quantum optimization algorithm. We describe a basic toolbox natural in trapped-ion quantum platforms, which can be used as building blocks to implement the method for complex Hamiltonian models. Finally, we numerically simulate the method for some interesting examples that could be implemented in state-of-the-art trapped-ion setups.

\acknowledgements 

This work was supported by the National Key Research and Development Program of China under Grants No. 2016YFA0301900 and No. 2016YFA0301901 and the National Natural Science Foundation of China Grants No. 11374178, No. 11574002, and No. 11504197. I. A. acknowledges support from Basque Government PhD Grant PRE-2015-1-0394, J. C. acknowledges support from Juan de la Cierva Grant IJCI-2016-29681, and L. L. acknowledges support from Ram\'{o}n y Cajal Grant RYC-2012-11391. We also acknowledge funding from Spanish MINECO/FEDER FIS2015-69983-P and Basque Government IT986-16.

\appendix

\section{Probabilistic Cooling Effect}

For an arbitrary input state $\ket{\psi}=\sum_jc_j\ket{j}$, the state for the system plus the ancilla qubit after applying $\hat W_\gamma(\tau)$ is denoted as $\ket{\Psi}\equiv\hat W_\gamma(\tau)\ket{\psi}\ket{0_{\rm A}}$. The values of the average energy with respect to the input state and conditioned output states are
\begin{eqnarray}
\average{\hat H}&=&\expect{\hat H_{\rm S}}{\psi}=\sum_jE_j\left|c_j\right|^2,\\
\average{\hat H}^{(0)}&=&\expect{\hat H_{\rm S}\otimes|0_{\rm A}\rangle\langle0_{\rm A}|}{\Psi}\nonumber\\
&=&\frac{1}{P_0}\sum_jE_j\left|c_j\right|^2\cos^2\left[\left(E_j+\gamma\right)\tau\right],\nonumber\\
\average{\hat H}^{(1)}&=&\expect{\hat H_{\rm S} \otimes |1_{\rm A}\rangle\langle1_{\rm A}| }{\Psi}\nonumber\\
&=&\frac{1}{P_1}\sum_jE_j\left|c_j\right|^2\sin^2\left[\left(E_j+\gamma\right)\tau\right],\nonumber
\end{eqnarray}
where $P_{0} =\sum_j |c_j|^2 \cos^2\left[\left(E_j+\gamma\right)\tau\right]$ and $P_{1}=\sum_j|c_j|^2\sin^2\left[\left(E_j+\gamma\right)\tau\right]$.
For small $\tau$, i.e. $(E_j+\gamma)\tau \ll 1$, we expand $\average{\hat H}^{(0)}$ to the second order of $\tau$, and calculate the difference between $\average{\hat H}^{(0)}$ and $\average{\hat H}$ as follows,
\begin{eqnarray}\label{eq:quasi_cooling}
\average{\hat H_{\rm S}}-\average{\hat H_{\rm S}}^{(0)}=\sum_{j}(|c_j|^2-|c'_j|^2)E_j,
\end{eqnarray}
where the new probability distribution is related with the old one as
\begin{eqnarray}\label{eq:quasi_cooling2}
\frac{c'_j}{c_j}&=&\frac{1}{P_{0}} \cos{[(E_j+\gamma)\tau]} \\ 
&\approx&\frac{1}{P_{0}}(1-\frac{1}{2}(E_j+\gamma)^2\tau^2)+{\mathcal O}\left(\tau^4\right). \nonumber
\end{eqnarray}
As $E_0+\gamma\geq0$, the factor $(1-\frac{1}{2}(E_j+\gamma)^2\tau^2)$ decreases for higher excited states, meaning that the higher the energy of the state, the lower the probability amplitude at the output state.
Similarly, we can calculate the difference between $\average{\hat H_{\rm S}}^{(1)}$ and $\average{\hat H_{\rm S}}$ to the leading order and obtain the following expression,
\begin{eqnarray}
\average{\hat H_{\rm S}}^{(1)}-\average{\hat H_{\rm S}}=\sum_{j}(|c''_j|^2-|c_j|^2)E_j,
\end{eqnarray}
where now the new probability distribution is related with the old one as
\begin{eqnarray}
\frac{c''_j}{c_j}&=&\frac{1}{P_{1}} \sin{[(E_j+\gamma)\tau]} \\ 
&\approx&\frac{1}{P_{1}}(E_j+\gamma)\tau+{\mathcal O}\left(\tau^3\right). \nonumber
\end{eqnarray}
Notice that in this case the lower the energy of the state, the lower the ratio between the new and the old probability amplitude.


\begin{thebibliography}{99}
\bibitem{Bounded} In our case, we use the Brent method to find a local minimum in the interval $0 < \tau_k < 1$.
\end{thebibliography}
\end{document}